# Brillouin amplification in phase coherent transfer of optical frequencies over 480 km fiber

O. Terra[1,2], G. Grosche and H. Schnatz

*Physikalisch- Technische Bundesanstalt, Bundesallee 100, 38116 Braunschweig, Germany*
[1] *Osama.terra@ptb.de*

**Abstract:** We describe the use of fiber Brillouin amplification (FBA) for the coherent transmission of optical frequencies over a 480 km long optical fiber link. FBA uses the transmission fiber itself for efficient, bi-directional coherent amplification of weak signals with pump powers around 30 mW. In a test setup we measured the gain and the achievable signal-to-noise ratio (SNR) of FBA and compared it to that of the widely used uni-directional Erbium doped fiber amplifiers (EDFA) and to our recently built bi-directional EDFA. We measured also the phase noise introduced by the FBA and used a new and simple technique to stabilize the frequency of the FBA pump laser. We then transferred a stabilized laser frequency over a wide area network with a total fiber length of 480 km using only one intermediate FBA station. After compensating the noise induced by the fiber, the frequency is delivered to the user end with an uncertainty below $2\times10^{-18}$ and an instability $\sigma_y(\tau) = 2\times10^{-14}/(\tau/s)$.

## References and Links

1. S. Diddams J. Bergquist, S. Jefferts, C. Oates, "*Standards of time and frequency at the outset of the 21$^{st}$ century,*" Science **306**, 1318 (2004).
2. L. Ma, P. Jungner, J.Ye, and J. Hall, "*Delivering the same optical frequency at two places: accurate cancellation of phase noise introduced by an optical fiber or other time-varying path*," Opt. Lett. **19**, 1777 (1994)
3. J. H. Jiang, F. Kéfélian, S. Crane, O. Lopez, M. Lours, J. Millo, D. Holleville, P. Lemonde, Ch. Chardonnet, A. Amy-Klein, and G. Santarelli, "*Transfer of an optical frequency over an urban fiber link,*" J. Opt. Soc. Am. B, **25**, 2029 (2008).
4. O. Terra, G. Grosche, K. Predehl, R. Holzwarth, T. Legero, U. Sterr, B. Lipphardt, and H. Schnatz, "*Phase-coherent comparison of two optical frequency standards over 146 km using a telecommunication fiber link*," Appl. Phys. B **97**, 541 (2009).
5. G. Grosche, O. Terra, K. Predehl, R. Holzwarth, B. Lipphardt, F. Vogt, U. Sterr, and H. Schnatz, "*Optical frequency transfer via 146 km fiber link with $10^{-19}$ relative accuracy,*" Opt. Lett. **34,** 2270 (2009).
6. E. Desurvire, "*Erbium-doped fiber amplifiers: principle and applications*," Wiley-Interscience publication, 1994.
7. H. Schnatz, O. Terra, K. Predehl, T. Feldmann, T. Legero,B. Lipphardt, U. Sterr, G. Grosche, R. Holzwarth, T. W. Hänsch,.T. Udem, Z. H. Lu, L. J. Wang, W. Ertmer, J. Friebe, A. Pape, E.-M. Rasel, M. Riedmann, and T. Wübbena, "*Phase-Coherent Frequency Comparison of Optical Clocks Using a Telecommunication Fiber Link*", IEEE Transactions on Ultrasonics, Ferroelectrics, and Frequency Control **57**, 175 (2010).
8. K. Predehl, R. Holzwarth, T. Udem, T. W. Hänsch, O. Terra, G. Grosche, B. Lipphardt, and H. Schnatz, "*Ultra Precise Frequency Dissemination across Germany - Towards a 900 km Optical Fiber Link from PTB to MPQ*," in Conference on Lasers and Electro-Optics/International Quantum Electronics Conference, OSA Technical Digest (CD) (Optical Society of America, 2009), paper CTuS2.
9. N. Olsson, J. van der Ziel, "*Cancellation of fiber loss by semi-conductor laser pumped Brillouin amplification at 1.5 µm,*" Appl. Phys. Lett. **48**, 1329 (1986).
10. R.Tkach, A.Chraplyvy, "*Fibre Brillouin amplifiers,*" Optical and Quantum Electronics **21,** S105 (1989).
11. M. Ferreira, J. Rocha, J. Pinto, "*Analysis of the gain and noise characteristics of fibre Brillouin amplifiers,*" Optical and Quantum Electronics **26**, 35 (1994).
12. R. Smith, "*Optical Power Handling Capacity of Low Loss Optical Fibers as determined by Stimulated Raman and Brillouin Scattering,*" Appl. Opt. **11**, 2489 (1972) .
13. E.Ippen, R. Stolen, "*Stimulated Brillouin scattering in optical fibers,*" Appl. Phys. Lett. **21**, 539 (1972).
14. G.Agrawal, "*Applications of Nonlinear Fiber Optics,*" Academic Press (2001).
15. G. Grosche, B. Lipphardt, H. Schnatz, "*Optical frequency synthesis and measurement using fiber-based femtosecond lasers,*" Eur. Phys. J. D **48**, 27 (2008).
16. J. Geng, S. Staines, M. Blake, and S. Jiang, "*Distributed fiber temperature and strain sensor using coherent radio-frequency detection of spontaneous Brillouin scattering*," Appl. Opt. **46**, 5928 (2007).

17. F. Walls, A. Clements, C. Felton, M. Lombardi, and M. Vanek, "*Extending the Range and Accuracy of Phase Noise Measurements,*" National Institute of Standards and Technology (NIST) Technical Note **1337,** TN129 (1990).
18. Williams, W. Swann, N. Newbury, "*High-stability transfer of an optical frequency over long fiber-optic links,*" J. Opt. Soc. Am. B **25**, 1284 (2008).
19. E. Rubiola, "*On the measurement of frequency and of its sample variance with high-resolution counters,*" Review of Scientific Instruments **76**, 054703 (2005).
20. W. Lee, D. Yu, C. Park, J. Mun, "*The uncertainty associated with the weigted mean frequency of a phase-stabilized signal with white phase noise,*" Metrologia **47,** 24 (2010).

## 1. Introduction

The development of optical frequency standards (optical clocks), will eventually lead to a new definition of the second [1]. One perquisite for a future redefinition is a suitable method to disseminate such ultra- stable optical frequencies over long distances. Optical fibers are particularly suited to transfer such frequencies to a distant user, provided that the attenuation of the signal and frequency fluctuation introduced by the acoustic and thermal fluctuations on the fiber are compensated. Such a compensation scheme requires to reflect part of the light back to the transmitting station to form a phase detection interferometer, which means that part of the light has to travel twice the distance to the destination [2]. In optical frequency transfer applications, Erbium doped fiber amplifiers (EDFA) are currently used to amplify the signal coherently. Typically, two uni-directional EDFA are used to amplify light in each direction [3]. The problem of using two separate EDFA is not only that the system will become more complex but also that the assumption that the forward and the return directions have the same noise will break down. Thus uncompensated residual noise will increase. To avoid this, we have recently built a bi-directional EDFA using the same gain medium for amplification of light in both directions [4, 5]. However, we have observed saturation of the gain medium and lasing effects caused mainly by Rayleigh scattering [6]. As a consequence the gain of the bi-directional EDFA has to be kept below 25 dB. In an amplifier chain, the signal-to-noise ratio of the EDFA is degraded due to amplified spontaneous emission (ASE) when the power of the input signal drops below a certain limit. Therefore, the distance between two bi-directional EDFAs is limited to less than 120 km. In consequence, the number of intermediate amplifier stations in an envisaged 900 km fiber link in Germany [7, 8] will increase, as well as the effort of controlling and maintaining these stations.

In order to keep the number of intermediate stations as low as possible, we have investigated the use of fiber Brillouin amplification (FBA) [9, 10, 11] as an alternative technique. This technique enables the amplification of a very small input signal (a few nano Watt) by more than 50 dB in a single gain step, with relatively low pump powers (about 30 mW). Moreover, it enables bi-directional amplification because it uses the fiber itself as the gain medium and different sections of the same fiber for each direction.

In section 2 we give a brief overview about the main properties of FBA, describe a test setup used to compare the amplification and signal-to-noise ratio (SNR) of FBA with that of our home-made bi-directional EDFA. We describe a measurement of the phase noise introduced by FBA and a new and simple method to stabilize the pump frequency for FBA, using the DC power of the stimulated Brillouin scattering (SBS) signal. In section 3 we demonstrate, for the first time, an optical frequency transfer over 480 km fiber link of a wide area network using FBA. In this case different parts of this fiber link act as gain medium for the FBA process and pump lasers were installed at the local and user ends, and at one intermediate station. The instability, the accuracy and the residual phase noise of the transmitted optical frequency are presented.

## 2. Fiber Brillouin amplification (FBA)

Stimulated Brillouin scattering (SBS) is a nonlinear process which results from the interaction of light with stimulated acoustic waves. In fused silica single mode fibers, acoustic waves with velocity $v_A$ ($v_A \approx 6 \times 10^3$ m/s) back-scatter light and shift down its frequency by ($v_B = 2 n v_A / \lambda$). This shift frequency is about 11 GHz for refractive index n = 1.451, and a wavelength of $\lambda = 1.5$ μm. The threshold power for SBS is given by [12, 13, 14]:

$$P_{crit} = \frac{21 A}{\gamma L_{eff}} (1 + \frac{\Delta v_{laser}}{\Delta v_B}) \qquad (1)$$

where A is the effective mode area of the fiber ($1 \times 10^{-10}$ m$^2$), and $\gamma$ is the gain coefficient of the nonlinear process ($5 \times 10^{-11}$ m/W). The effective gain length is $L_{eff} = (1-e^{\alpha L})/\alpha$, therefore $L_{eff} \approx 21$ km for an attenuation coefficient of $\alpha = 0.2$ dB/km. The Brillouin linewidth depends on the life time of the scattered excitation ($\tau_B$), $\Delta v_B = 1/\pi\tau_B$ Hz. For a 148 km long fiber we measured an SBS linewidth of 10 MHz. We used a laser with linewidth of about $\Delta v_{laser} = 5$ kHz, which is much narrower than $\Delta v_B$. Therefore, the SBS threshold power is only 1 mW according to equation (1).

If a pump laser with a power ($P_{pump}$) is injected in the fiber, it will cause SBS with a gain of [9]

$$g = \frac{\gamma L_{eff} P_{pump}}{A} \qquad (2)$$

If a signal is injected in opposite direction to the pump laser and with a frequency shifted down by the SBS frequency ($v_B$), it will be amplified by a process called fiber Brillouin amplification (FBA) [10, 11]. Due to the high gain of the Brillouin process (9 dB /mW) (Equation 2), very weak signals can be amplified with a very high gain using only few mW of pump power.

### 2.1 Amplification and signal-to-noise ratio

For measuring the amplification due to FBA we performed a simple test. We used a 25 km spool of single mode fiber (SMF28) as FBA gain medium and narrow linewidth fiber lasers as pump and signal lasers. These lasers have a wavelength of $\lambda = 1542$ nm, a linewidth of about 5 kHz, up to 100 mW output power and a tuning range of about 2 nm. As described in [15], the signal laser is stabilized to our optical frequency reference to reach a linewidth of about 10 Hz.

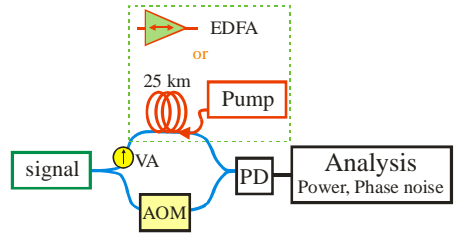

Fig. 1: Set-up to measure the amplification of FBA or EDFA: VA: variable attenuator, PD: photodetector, signal: signal laser with isolator, pump: pump laser, AOM: acousto-optic modulator, a 25 km fiber is used as gain medium for FBA.

With our 10 GHz resolution optical spectrum analyzer it is not possible to resolve the amplification caused by FBA, because the SBS width itself is only about 10 MHz. Therefore, we constructed a Mach-Zehnder interferometer with the 25 km fiber spool in the measurement arm and used the other arm as reference, as shown in fig. (1). A variable attenuator is installed before the 25 km arm to adjust the input power for the 25 km fiber without changing the power in the reference arm. The pump laser is injected from the opposite end of the signal with frequency shifted $\nu_B = 10.972$ GHz from the signal. We tuned the variable attenuator at the input of the 25 km fiber to produce different output powers, to simulate different fiber lengths. A photo detector is used to detect the heterodyne beat signal between the amplified light after the 25 km fiber and the light from the reference arm. This allows to measure the gain and the SNR of the device with a RF spectrum analyzer.

For comparison we performed the same measurement for a bi-directional EDFA and a commercial single-pass EDFA and include the result in fig. (2). Note, the latter is not usable in the optical frequency transfer scheme, since bi-directional operation is required to compensate the fiber phase noise.

Fig. 2 shows the gain (a) and the SNR (b) for different signal powers for the devices under test. Although both the bi-directional EDFA and FBA achieve almost the same SNR, the gain of FBA is significantly higher than that of the bi-directional EDFA. This is especially true for low signal powers. As an example, the gain of FBA is about 1000 times higher than that of the bi-directional EDFA for signal powers less than 50 nW. This level corresponds to the typical signal power after a distance of $>250$ km (input power $= 5$ mW, attenuation $= 0.2$ dB/km, wavelength $= 1.54$ µm). While the bi-directional EDFA thus only allows a span up to 120 km, FBA allows a span of approximately 250 km between intermediate amplification stations.

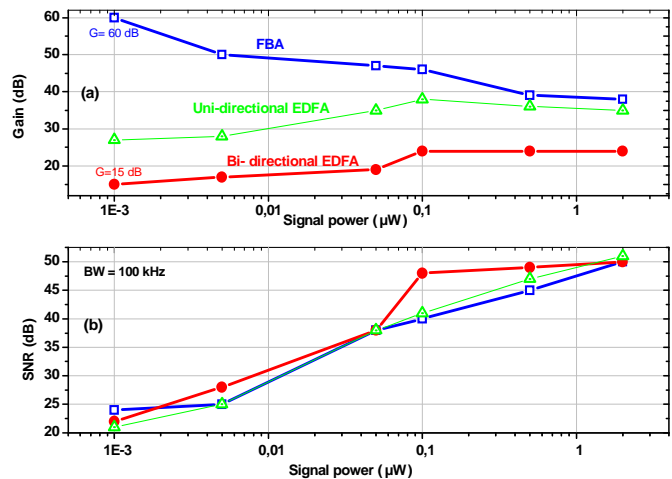

Fig. 2: FBA in comparison to EDFAs (uni-directional and bi-directional) for different signal powers received at the ouput of a 25 km fiber: (a) gain (b) SNR. The spectrum analyzer bandwidth is 100 kHz.

This opens two possibilities: using an additional fiber spool as discrete FBA-module is a suitable solution when operational or safety requirements forbid injecting more than a few mW optical power into the transmission fiber. However, when higher power levels are allowed in the installed transmission fiber, this fiber itself can be used as gain medium. In order to test this option of a distributed amplifier further, we performed measurements using FBA and the bi-directional EDFA on 148 km and 332 km of installed, commercial fibers. The installed dark fiber is buried and part of a wide-area network connecting PTB with other research institutes in Germany [7].

The 148 km fiber link consists of two 74 km fibers connecting Physikalisch-Technische Bundesanstalt (PTB) at Braunschweig to Institute of quantum optics (IQ) at Hanover. The 332 km link consists of two 166 km fibers which go from PTB, Braunschweig, south-east to a network node near Coermigk/Halle as shown in fig. (6). Replacing the 25 km fiber spool by either the 148 km link or the 332 km link, the setup in fig. (1) is used to make the measurements shown in fig. (3). The measurements shows that for the 148 km link, intrinsic FBA achieves 18 dB more gain than our bi-directional EDFA. For the 332 km link, the difference is even larger: FBA gives 45 dB more gain than the EDFA. This result shows that FBA is more suitable for use in long fiber links. The Brillouin shift frequency is measured to be 10.974 GHz and 10.970 GHz for the 148 km and the 332 km fiber link, respectively, with only 4 MHz difference.

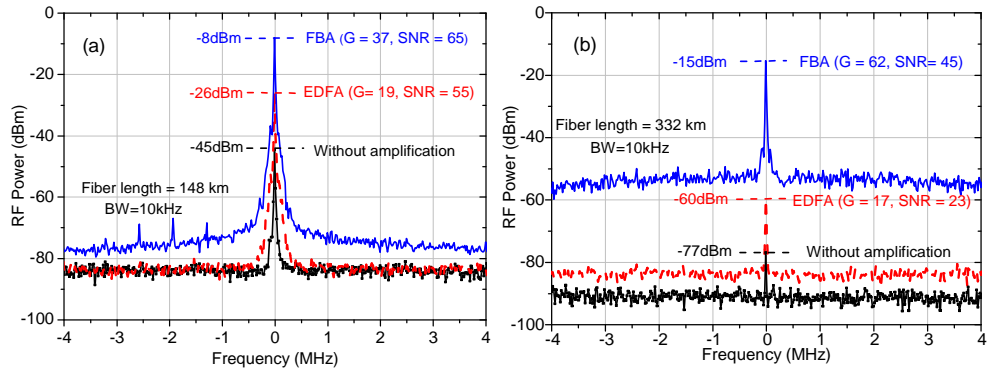

Fig. 3: Detected heterodyne beat power (RF) obtained with FBA and with EDFA, when 5 mW signal power are injected into (a) 148 km fiber ($P_{pump}$ = 20 mW, $\nu_B$ = 10.974 GHz) and (b) 332 km fiber ($P_{pump}$ = 40 mW, $\nu_B$ = 10.970 GHz).

*2.2 Phase noise*

One of the most important aspects in frequency transfer applications is how much additional phase noise is introduced by the amplifier to the signal propagating through the fiber. We used the Mach-Zehnder interferometer shown in fig. (1) to measure the phase noise of the bi-directional EDFA and FBA. Fig. (4a) shows the measured phase noise for the free-running interferometer without (black line) and with installed EDFA (red circles o). At low Fourier frequencies ($f$ < 30 Hz) the EDFA possibly adds a small amount of phase noise – this is still less than 0.1 rad$^2$/Hz at $f$ = 1 Hz. We attribute this phase noise to the approximately 6 m fiber used as gain medium inside the amplifier.

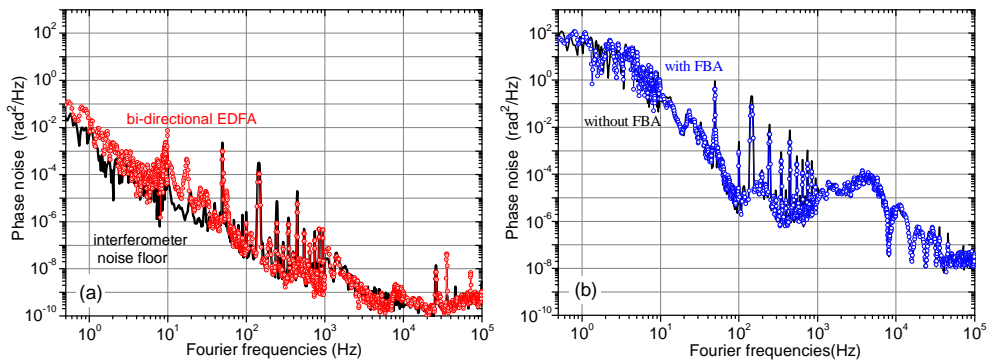

Fig. 4: (a) Phase noise of the free-running interferometer without (black line) and with a bi-directional EDFA (red o). (b) Phase noise of 25 km spooled SMF28 fiber without (black line) and with using FBA (blue o).

To measure the phase noise introduced by FBA, a 25 km fiber spool is used as FBA gain medium. Such a spool is quite sensitive to environmental perturbations, which results in phase noise. To reduce this effect, we put the spool inside an acoustic isolation box. Still, the measured phase noise for the unstabilized 25 km fiber (black line in Fig. 4(b)) exceeds one radian at frequencies below 10 Hz. With the pump laser turned on, we then measured the phase noise of FBA (blue circles o). The phase noise due to the FBA process is less than that of the 25 km fiber alone, since both curves coincide. The total phase noise is less than that encountered on long fiber links. It can therefore be compensated with the standard technique [2, 5] if we assume that light traveling in either direction will experience the same phase noise. An experiment over a 480 km link demonstrating this is described in section 3.

*2.3 Pump laser stabilization*

In order to keep the peak of the SBS gain spectrum matched to the frequency of the signal, the pump frequency needs to be stabilized relative to the signal frequency. This means that the frequency difference between both lasers equals the Brillouin scattering frequency $\nu_B$ ($\nu_B$ = 10.970 GHz for the 148 km fiber). If pump and signal lasers are at the same site, a beat between the signal and the pump laser can be detected with a fast photodetector (bandwidth > 11 GHz). The beat signal is then locked to a microwave reference. But this technique needs complex and expensive equipment, which handles GHz frequencies. Furthermore, this technique will not be applicable if the pump laser is located at an intermediate station.

We have developed an simpler method based on the observation that the SBS power (DC) increases when it's frequency matches the signal frequency. Although the signal power is less than 1 µW after the 148 km fiber, Fig. (5.a) shows the SBS power change when the signal frequency is swept over the SBS profile. We used this change as a discriminator to lock the pump laser at the peak of the curve which corresponds to the maximum amplification. A small modulation of the pump laser frequency can be used to lock the pump laser to the maximum of the SBS, using a lock-in amplifier.

To determine the SBS center frequency change with temperature [16], we measured the temperature dependence of the 25 km fiber in the range from 16 °C to 25 °C. The SBS center frequency temperature dependence is found to be 0.4 MHz/K. Typical peak to peak diurnal temperature variations for a buried fiber (depth < 1.5 m ) are well below 1 K. Even if we assume an upper limit for temperature variations of about 10 K, the stabilization circuit can easily handle a pump laser frequency shift of about 4 MHz.

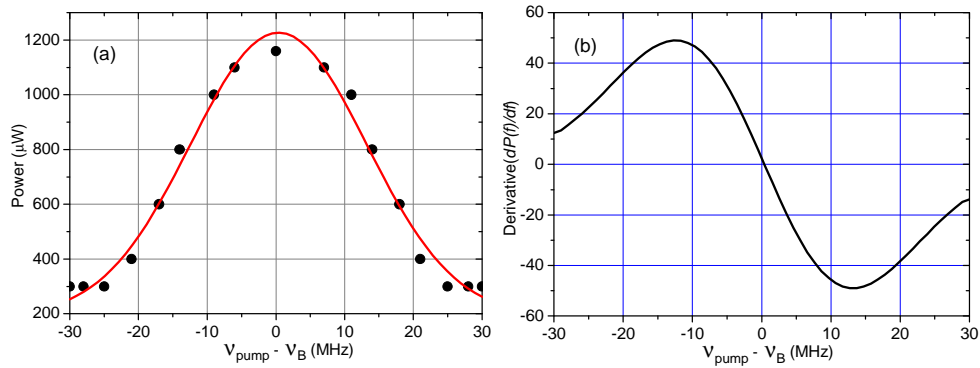

Fig. 5: (a) SBS power change when the signal frequency is swept around the SBS gain curve and a Gaussian fit (signal power is less than 1 µW after 148 km fiber) (b) Derivative of the SBS signal power.

## 3. Frequency transfer over a 480 km optical fiber

To transfer a stable optical frequency from one location to a remote user over an optical fiber link phase noise accumulated along the fiber as well as attenuation have to be compensated. In this section we describe the transfer of an optical carrier over 480 km using the fiber stabilization scheme described in [4, 5] and an amplification scheme that utilizes FBAs at each end of a 480 km fiber link and in one intermediate station only.

The 480 km fiber consists of four commercial fibers with overall attenuation of 115 dB, two of them are connecting PTB, Braunschweig to Cörmigk with attenuation of about 70 dB and the other two connecting PTB to the Institute of Quantum Optics (IQ), Hanover with attenuation of about 45 dB. In this setup local and user end as well as the intermediate station are located at PTB, which allows easy testing. The light is sent in the first fiber (166 km) to Cörmigk, where it is returned to the intermediate station at PTB by fiber 2. The light is amplified and sent in the third fiber (74 km) to IQ, where it is connected to a forth fiber to return the light back to PTB (the user end).

A transfer laser is locked to an optical frequency standard with the help of a femtosecond frequency comb. The transfer laser is used to transfer the stability of the frequency standard from the local to the user end, where it is used to measure the frequency of another laser or may be used for other applications. As acoustic and thermal fluctuations on the fiber link induce phase fluctuations on the light propagating in the fiber, a part of the light is reflected using a Faraday mirror at the user end back to the local end. A beat between the reflected light and a reference arm detects the phase noise introduced by the fiber link. A phase-locked loop (PLL) is then used to control an acoustic-optic-modulator (AOM1) to compensate this phase noise. AOM2 is used to discriminate light reflected at the user end by shifting its frequency from back scattered light in the fiber and back reflected light at the local end. The compensation system is shown in fig. (6).

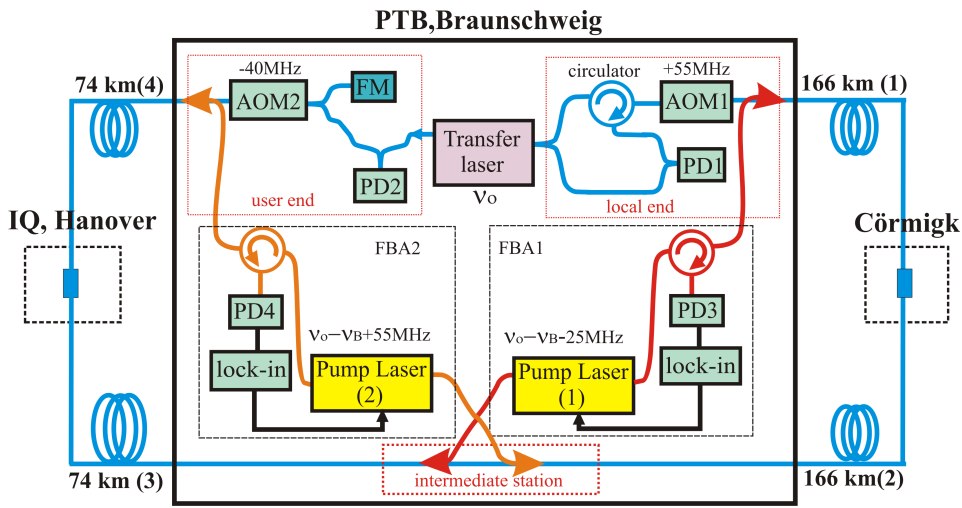

Fig. 6: Brillouin amplification in a frequency transfer system, AOM: Acousto-optic modulator, PD: photodetector, FM: Faraday mirror (for details see text).

For a full round-trip in the fiber the signal accumulates 230 dB loss before it reaches the phase compensation detector PD1. In order to compensate this loss, we used two FBA amplifiers for pumping four fiber sections. FBA amplifier is used at the intermediate station to amplify the light in the return direction with pump power of 45 mW and in the forward direction with pump power of 30 mW. It is also used at the local end with pump power of 9 mW to amplify the return light and at the virtual user end with pump power of 18 mW to amplify the forward light. We used one amplifier for each direction because of the 80 MHz frequency difference between the forward and the return light introduced by the AOM. For the forward direction FBA2 is used with frequency $v_{FPump} = v_o - v_B + 55 MHz$ and for the return direction FBA1 is used with frequency $v_{RPump} = v_o - v_B - 25 MHz$. Where $v_B$ is the Brillouin frequency and $v_o$ is the frequency of the transfer laser (signal). A part of the Brillouin-scattered light is directed by circulators to the DC photodetectors PD (3 and 4), where it is used to stabilize the frequency of the pump laser using the method discussed in section (2.3).

A beat at the user end between the forward light and the light from a reference arm (out-of-loop beat) indicates the 480 km fiber link stability. The signal is analyzed in the frequency domain with a FFT spectrum analyzer; in the time domain we use a Π-type frequency counter with zero dead time.

The phase noise of the out-of-loop beat is measured before and after applying the compensation scheme by detecting the phase changes with respect to a reference oscillator using a digital phase detector, after division by a suitable ratio to keep the phase changes below one radian [17]. Fig. (7) shows the phase noise in (rad$^2$/Hz) for the out-of-loop signal measured before and after applying the compensation. The phase noise reduction is 39 dB at 1 Hz, which is near to the theoretical limit of 41 dB at 1 Hz predicted by Williams et al. [18], $S_D(f) = (1/3)(2\pi)^2 \tau_{delay}^2 f^2 S_\phi(f) = 7.5 \times 10^{-5} f^2 S_\phi(f)$, where $S_\phi(f)$ is the phase noise of the free-running fiber and $f$ is the Fourier frequency. It is close to optimal (green o in fig. 7) up to the compensation bandwidth of about 100 Hz. The compensation bandwidth is $1/4\tau_{delay}$, where $\tau_{delay}$ is the time for a single trip time in the fiber.

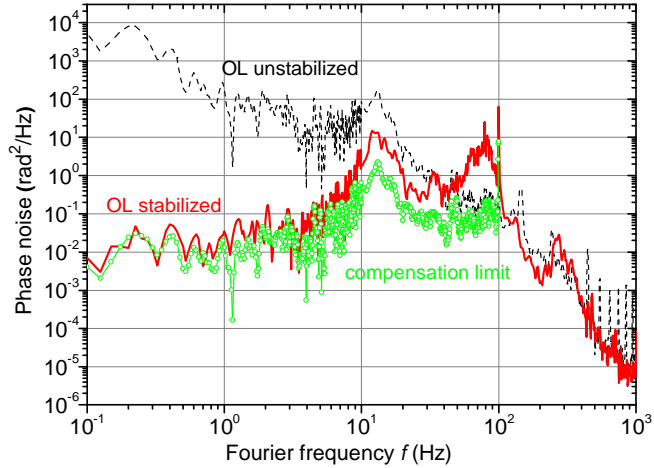

Fig. 7: Phase noise of the out-of-loop signal (OL) before (black dashed), and after (red solid) compensation. The green curve (o) gives the theoretical compensation limit according to [19].

We used a Π-type frequency counter with 1 second gate time to measure the frequency in the time domain and used the Allan deviation (ADEV) as a statistical measure of the frequency stability [19]. Fig. (8) shows of the ADEV of the out-of-loop signal before (black ■) and after (red o) the phase compensation. After applying the compensation, the stability of the transferred frequency reaches $\sigma_y(\tau) = 2\times 10^{-14}/(\tau/s)$. This value is in good agreement with the value calculated from the phase noise (see Fig. 7). The ADEV curve follows a $1/\tau$ slope and reaches a value of $2 \times 10^{-18}$ after about 2 hours. Comparing the achieved instability for this 480 km stabilized link with our previous measurements over 146 km [4], we observe excellent agreement with the scaling law ($\sigma_y \sim L^{3/2}$) derived in [18]. Thus the new measurements support our prediction [5] that we can achieve an instability of $\sim 5\times10^{-14}/(\tau s^{-1})$ for a 900 km link connecting PTB Braunschweig and MPQ Garching.

The mean value of the transmitted frequency is shifted from that of the reference laser by 64 μHz with a statistical uncertainty of 54 μHz. The statistical uncertainty is obtained by dividing the standard deviation of the measurement by $N$ since it is white phase noise and not by $\sqrt{N}$, as discussed in [20], where $N = 65000$ is the total number of data points.

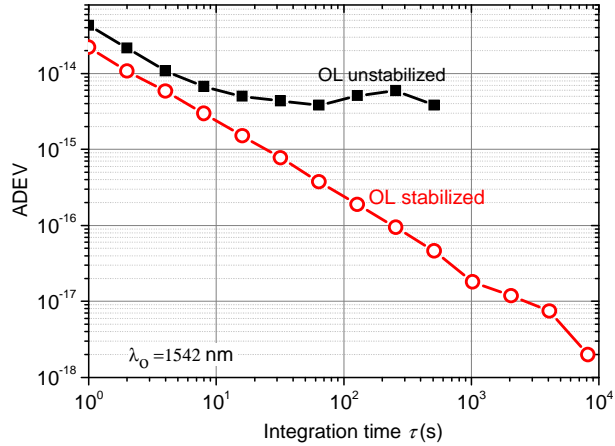

Fig. 8: Out-of-loop signal (OL) with compensated (red o) and uncompensated phase noise (black ■).

## 4. Conclusion

We have presented a fiber Brillouin amplifier (FBA) as an alternative to the currently used EDFA amplifier in frequency transmission applications. FBA offers bi-directional amplification with a SNR ratio comparable to that of the currently used EDFA and with high gain of about 50 dB, specially when the signal is very small (several nW). FBA phase noise was measured, and a simple and new method for stabilizing FBA pump laser was discussed. Finally, FBAs at the user and local ends and at one intermediate station only, were used to transfer an optical carrier frequency over 480 km fiber. The relative instability of the fiber link is $\sigma_y(\tau) = 2\times10^{-14}/(\tau/s)$ and reaches $2\times10^{-18}$ after about two hours. Thus, the link will not limit a frequency comparison between today's best optical clocks. Moreover, this is to our knowledge the largest distance bridged using only one intermediate amplifier station and with the lowest instability reported so far.

**Acknowledgment**

The authors would like to thank Thomas Legero, Katharina Predehl and all the staff members of PTB who have contributed to the results discussed in this report. The work was partly supported by DFG through the Centre for Quantum Engineering and Space-Time Research, QUEST. Osama Terra is supported by a scholarship from the Egyptian National institute of standards (NIS) and is a member of the Braunschweig International Graduate School of Metrology, IGSM.

[2]Home address: National Institute of Standards (NIS), Tersa st., Haram, Giza, Egypt.
P.O.Box: 136 Giza, Postal code:12211
e-mail: osama.terra@nis.sci.eg